\documentclass[twocolumn]{aastex631}

\shorttitle{Eccentricity Distribution Beyond the Snow Line}
\shortauthors{Stephen R. Kane \& Robert A. Wittenmyer}

\begin{document}

\title{Eccentricity Distribution Beyond the Snow Line and Implications
  for Planetary Habitability}

\author[0000-0002-7084-0529]{Stephen R. Kane}
\affiliation{Department of Earth and Planetary Sciences, University of
  California, Riverside, CA 92521, USA}
\email{skane@ucr.edu}

\author[0000-0001-9957-9304]{Robert A. Wittenmyer}
\affiliation{University of Southern Queensland, Centre for
  Astrophysics, West Street, Toowoomba, QLD 4350 Australia}


\begin{abstract}

A fundamental question in the study of planetary system demographics
is: how common is the solar system architecture? The primary
importance of this question lies in the potential of planetary systems
to create habitable environments, and dissecting the various
components of solar system evolution that contributed to a sustainable
temperate surface for Earth. One important factor in that respect is
volatile delivery to the inner system and the dependence on giant
planets beyond the snow line as scattering agents, particularly as
such cold giant planets are relatively rare. Here, we provide an
investigation of the eccentricity distribution for giant planet
populations both interior and exterior to their system snow lines. We
show that the median eccentricity for cold giants is 0.23, compared
with a far more circular orbital regime for inner planets. We further
present the results of a dynamical simulation that explores the
particle scattering potential for a Jupiter analog in comparison with
a Jupiter whose eccentricity matches that of the median cold giant
eccentricity. These simulations demonstrate that the capacity for such
an eccentric cold giant system to scatter volatiles interior to the
snow line is significantly increased compared with the Jupiter analog
case, resulting in a far greater volume of Earth-crossing
volatiles. Thus, many of the known systems with cold giant planets may
harbor water worlds interior to the snow line.

\end{abstract}



\section{Introduction}
\label{intro}

Surveys for exoplanet discovery have now been operating for several
decades, and have been increasing the sensitivity of such measurements
to a greater diversity of exoplanet masses and semi-major axes. For
example, radial velocity (RV) surveys have developed long
observational baselines \citep{fischer2016} that have probed into the
outer regions of planetary systems \citep{wittenmyer2013a}. With
thousands of exoplanets now known, the distribution of planetary
architectures is gradually becoming apparent
\citep{ford2014,winn2015,he2019,mishra2023a}. In particular, exoplanet
demographics may be compared to the architecture of the solar system
in order to place the masses and orbits of our planets within the
broader context of planetary formation and evolution
\citep{martin2015b,horner2020b,raymond2020a,kane2021d}. Such
comparisons have revealed that our system may be relatively unusual in
the harboring of several giant planets beyond the snow line, since the
prevalence of similar planets in other systems appears to be
relatively rare, even for solar-type stars
\citep{wittenmyer2011a,wittenmyer2016c,wittenmyer2020b,fulton2021,rosenthal2021,bonomo2023}. Since
Jupiter is the overwhelming dominant planetary mass within the solar
system, it is critically important that we understand the role of
giant planets in the evolution of terrestrial planets, especially
those that reside within the Habitable Zone (HZ) of the system
\citep{kasting1993a,kane2012a,kopparapu2013a,kopparapu2014,kane2016c,hill2018,hill2023}.

Giant planet formation within a protoplanetary disk can result in
numerous important planet-planet interactions that can play a dominant
role in the sculpting of the eventual planetary architecture
\citep{morbidelli2007b,raymond2008b,raymond2009b,kane2023a}, and can
limit the formation and stability of habitable planets
\citep{raymond2006a,kopparapu2010,kane2015b,kane2020b,kane2023c}. The
precise nature of how the presence of giant planets within a system
influences the formation of inner planets, the abundance of volatile
delivery, and the potential habitability of HZ planets is an active
area of research
\citep{morbidelli2000,morbidelli2016b,clement2022a}. A key
consideration is the location of the ``snow line'', which is the
radial distance from the center of a protostellar disk beyond which
volatiles (such as water) can efficiently condense to form ice
\citep{ida2005,kennedy2006b,kennedy2008a,kane2011d,ciesla2014}. As
giant planets form beyond the snow line, their accretion and migration
can lead to significant scattering of volatiles to the inner part of
the disk \citep{raymond2014d,raymond2017b,venturini2020b}. This
scattering effect is a crucial component for forming the volatile
inventory of terrestrial planets, most particularly the water mass
that can contribute to a sustained presence of surface liquid water
\citep{raymond2004a,ciesla2015b,marov2018,ogihara2023}. Moreover, the
eccentricity of planets beyond the snow line plays a role in the
scattering profile of encountered material, and the subsequent
prospects for impact scenarios with the inner planets. The solar
system giant planets have likely passed through eccentric phases of
their orbital evolution via planet-planet interactions and migration
processes, before arriving at their present near-circular orbits
\citep{clement2021a}. Exoplanets provide a statistical basis from
which to evaluate the eccentricity distribution
\citep{shen2008c,hogg2010,kane2012d,sagear2023} and the implications
for planet formation scenarios \citep{juric2008b,ida2013}. Given the
relative scarcity of their population, giant planets beyond the snow
line enable an exploration of the effect of their eccentricity
distribution on the disbursement of icy material throughout the
system.

Here, we present the results of a comparative study of the
eccentricity distribution of giant planets on either side of the snow
line, and the dynamical consequences for volatile delivery to the
inner region of planetary systems. Section~\ref{demo} describes the
selection criteria for our giant planet sample, and the eccentricity
distribution statistics with respect to the snow
line. Section~\ref{dynamics} provides the results of a dynamical
simulation that calculates the fractional amount of injected particles
that are scattered interior to the snow line, comparing a Jupiter
analog scenario with that of a Jupiter with considerably higher
eccentricity. Section~\ref{discussion} discusses the known occurrence
rate of cold Jupiters, and the implication of our dynamical results
for volatile delivery, planetary habitability, and the frequency of
water worlds. Finally, we provide concluding remarks and suggestions
for future work in Section~\ref{conclusions}.


\section{Demographics Beyond the Snow Line}
\label{demo}

Here, we describe our sample selection for exploring the eccentricity
distribution both interior and exterior to the snow line.


\subsection{Sample Selection and Snow Line}
\label{sample}

Our sample consists of known giant exoplanets detected beyond the snow
line of their host stars. The data for our sample were extracted from
the NASA Exoplanet Archive \citep{akeson2013} and are current as of
2023 October 20 \citep{nea}. We selected all planets with valid
(non-null) values for the semi-major axis ($a$), planet mass ($M_p$),
orbital eccentricity ($e$), and the mass of the host star
($M_\star$). This provided an initial sample of 1492 planets. We
further constrained the data by retaining only those planets whose
mass is equal to or greater than Saturn's mass ($\sim$0.3~$M_J$),
which produced a sample size of 846 planets. Finally, we calculated
the snow line for each of the host stars in the sample. The location
of the snow line was approximated as a function of stellar mass using
the relationship $a_{\mathrm{ice}} = 2.7 (M_\star / M_\odot)^2$,
derived by \citet{ida2005}, and which provides an estimated snow line
location of $\sim$2.7~AU for the solar case (see
Section~\ref{dynamics}). The sample of planets was divided into those
interior to the system snow line (651 planets) and those exterior (195
planets).


\subsection{Eccentricity Distribution}
\label{ecc}

\begin{figure*}
    \begin{center}
        \begin{tabular}{cc}
            \includegraphics[angle=270,width=8.5cm]{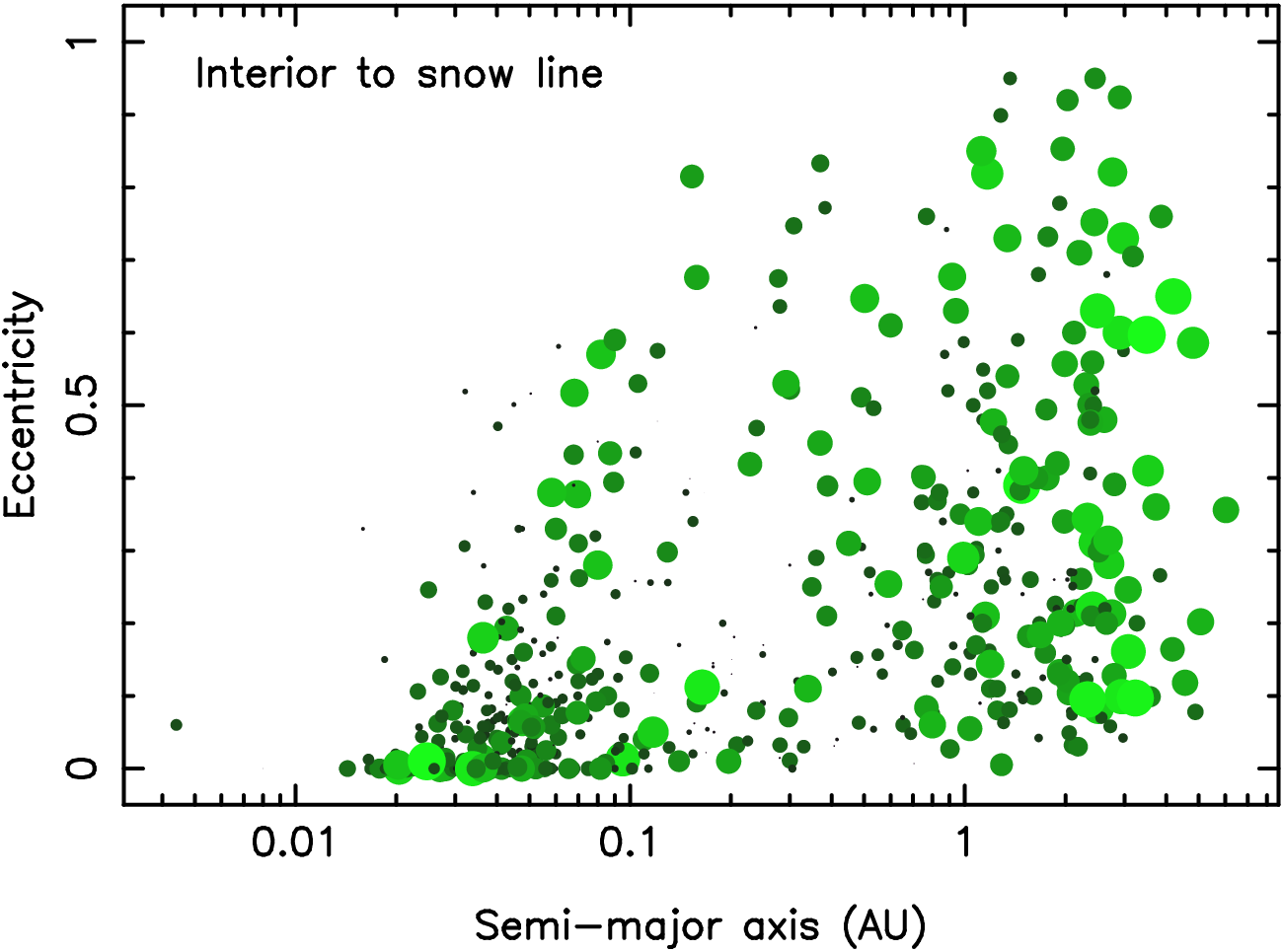} &
            \includegraphics[angle=270,width=8.5cm]{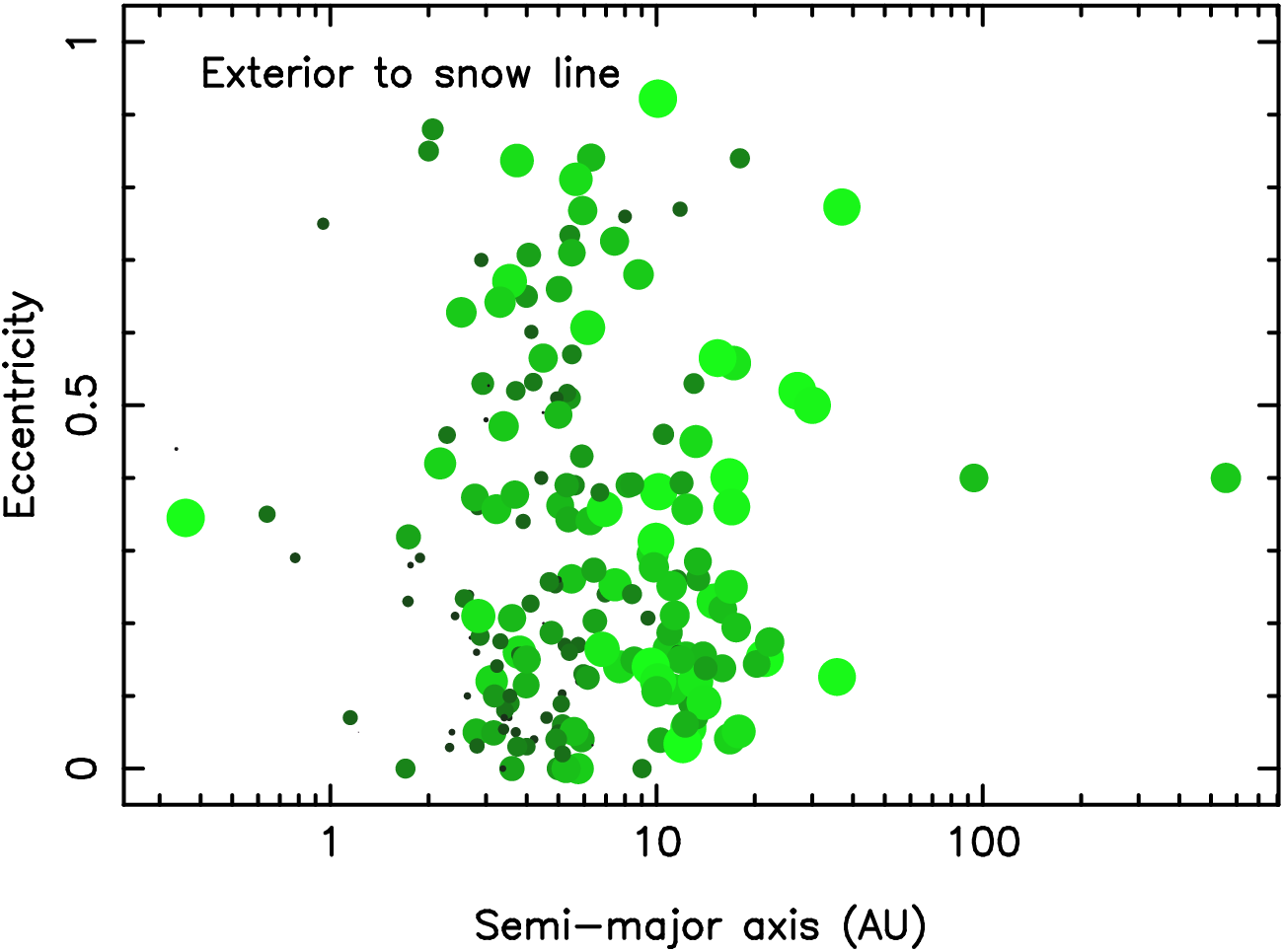} \\
            \includegraphics[angle=270,width=8.5cm]{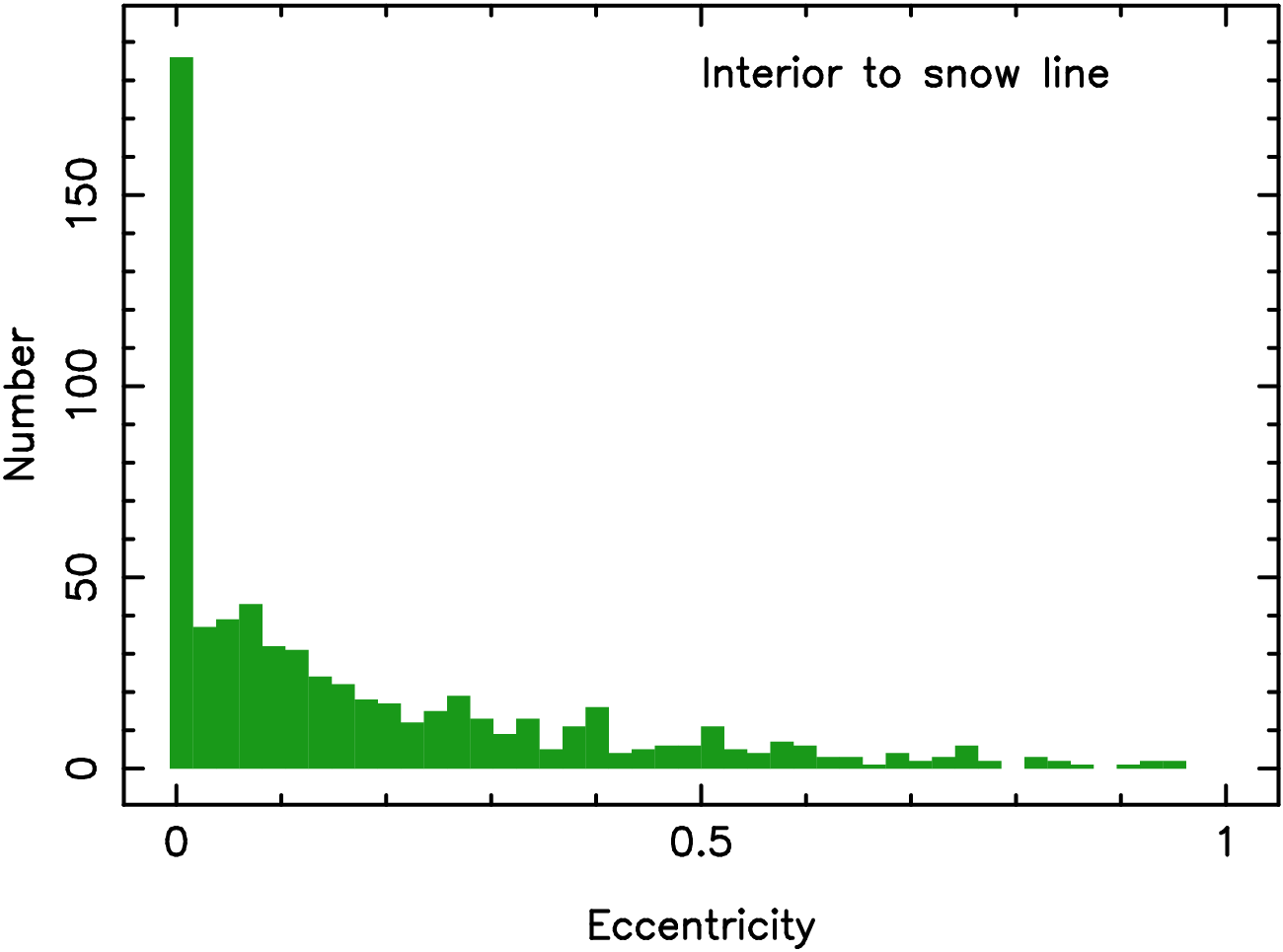} &
            \includegraphics[angle=270,width=8.5cm]{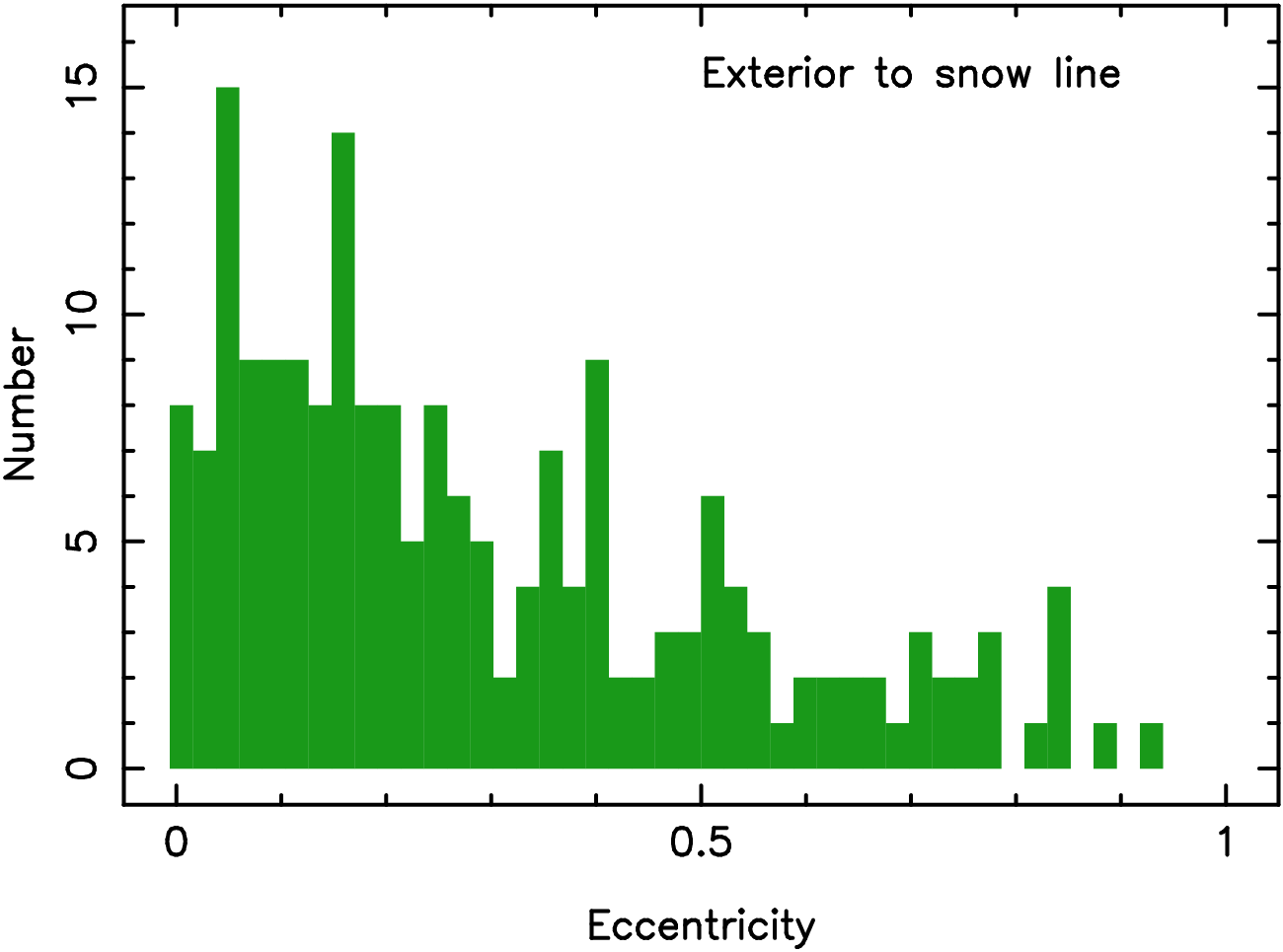}
        \end{tabular}
    \end{center}
  \caption{Distribution of eccentricities for all planets more massive
    than 0.3~$M_J$ interior to the snow line (left panels) and
    exterior to the snow line (right panels). For the top two panels,
    both the shade and the size of the plotted data are
    logarithmically proportional to the planet mass, where dark green
    indicates a low mass and light green indicates a high mass. The
    bottom two panels show the histograms of the eccentricity data
    appearing in the top two panels.}
  \label{fig:ecc}
\end{figure*}

Based on the sample selection described in Section\ref{sample}, we
examined the planetary property distributions as a function of
semi-major axis. The extracted eccentricity data are represented in
Figure~\ref{fig:ecc} as a function of semi-major axis, where the
planets interior to the snow line are shown in the left panels, and
those exterior to the snow line are shown in the right panels. For the
scatter plots shown in the top two panels, both the shade and size of
the data points are logarithmically proportional to the planet mass,
where dark green and light green indicate planets at the low and high
end of the mass range (see Section~\ref{sample}), respectively. Note
that, since many of the planets included in our sample are
non-transiting, many of the masses are minimum masses depending on
their orbital inclination relative to the plane of the sky. The data
plotted in the top-left panel show the well-known trend toward
circular orbits for small star-planet separations
\citep{juric2008b,shen2008c} that largely results from tidal
circularization \citep{goldreich1966a,ivanov2007a}. The top panels of
Figure~\ref{fig:ecc} also reveal a larger concentration of low mass
planets that lie interior to the snow line. For the planet population
interior to the snow line, the mean and median mass values are 2.74
and 1.28 Jupiter masses, respectively. For the planet population
exterior to the snow line, the mean and median mass values are 7.64
and 5.15 Jupiter masses, respectively. This difference in mass
distribution is in large part caused by the observational bias, both
in terms of the survey precision and duration requirements needed to
detect similar planets beyond the snow line
\citep{kane2007a,ford2008a,wittenmyer2013a}.

Of particular interest to our analysis is the eccentricity
distribution for those planets either side of the estimated snow line
in each system. It is apparent from Figure~\ref{fig:ecc} that the
eccentricities shown in the top-right panel are more evenly
distributed than those in the top-left panel. Quantitatively, for
those planets in our sample that are interior to the snow line, the
mean and median eccentricities are 0.18 and 0.01, respectively, with a
1$\sigma$ RMS scatter of 0.21. For those planets exterior to the snow
line, the mean and median eccentricities are 0.29 and 0.23,
respectively, with a 1$\sigma$ RMS scatter of 0.23. The large
difference between mean and median eccentricities for the planets
interior to the snow line suggests the sample is heavily skewed by
outliers, dominated by observational bias toward the detection of
inner planets and the tidal circularization that occurs within that
regime, and can be clearly seen in the histograms shown in the bottom
panels of Figure~\ref{fig:ecc}. By contrast, the proximity of the mean
and median for the sample beyond the snow line indicates a relatively
symmetrical distribution of the eccentricities. To quantify these
differences more thoroughly, we conducted a null hypothesis
Kolmogorov-Smirnov (K-S) test to assess the statistical significance
of the similarities between datasets. The K-S test produced a
probability of $\sim$0.0, thus rejecting the hypothesis that the two
samples are drawn from the same distribution. However, excluding
planets with an orbital period less than 10 days results in a median
eccentricity of 0.24 for planets interior to the snow line, more
closely resembling the eccentricity distribution exterior to the snow
line. A K-S test using this new criteria resulted in a probability of
$\sim$0.997, such that the hypothesis of the two samples being drawn
from the same distribution may be accepted. As a further test, we
considered the constraints of excluding planets with an orbital period
less than 10 days and also increasing the lower limit on planet mass
from 0.3~$M_J$. The K-S test using these revised samples produced a
probabilities of 0.646 and 0.327 for lower mass limits of 1.0~$M_J$
and 2.0~$M_J$, respectively. These results similarly lead to the
acceptance that both samples are drawn from the same distribution,
though at lower significance than the 0.3~$M_J$ lower mass
limit. These differences in K-S test outcome due to the choice of
lower mass limit are consistent with previous work that found an
eccentricity dependency on planet mass
\citep{jones2006,ribas2007,ford2008c,wright2009a}, which may be due to
an increase in planet-planet scattering events
\citep{raymond2010,ida2013}. Thus, our analysis indicates that the
differences between the two original samples described in
Section~\ref{sample} are largely due to the circularization of short
period planets, but also exhibit planet mass dependencies For our
subsequent dynamical simulations, we adopt the median eccentricity of
the sample beyond the snow line ($e = 0.23$) as a representative
example.


\section{Dynamical Consequences}
\label{dynamics}

An increase in eccentricity for a giant planet beyond the snow line
may have a significant effect on the planet's ability to scatter
material interior to the snow line. To test the effect of eccentricity
on scattering potential, we conducted a series of dynamical
simulations that explore the specific case of a Jupiter analog, and
the equivalent eccentricity case. We used the Mercury Integrator
Package \citep{chambers1999} with a hybrid symplectic/Bulirsch-Stoer
integrator with a Jacobi coordinate system
\citep{wisdom1991,wisdom2006b}, following the methodology described by
\citet{kane2014b,kane2019c,kane2021a,kane2023c}.

Two main simulation suites were conducted; one for the "Jupiter
analog" case that adopts the present Jupiter eccentricity of 0.049,
and one for the "eccentric Jupiter" case that adopts the median
eccentricity of giant planets beyond the snow line of 0.23 (see
Section~\ref{ecc}). A Jupiter mass and solar mass were adopted for the
planetary and star masses, respectively, and the semi-major axis of
the planet was set to 5.2~AU. The simulations within each suite
explored a location range of 3.0--8.0~AU in steps of 0.01~AU,
resulting in $\sim$500 simulations for each of the simulation
suites. The simulations at each semi-major axis step were run for
$10^5$ years with a time step of 10 days. At each semi-major axis
location, 100 particles, each with a mass of $10^{-6}$ Earth masses,
were injected into circular orbits at equally spaced starting
locations. At the conclusion of the simulation for a given semi-major
axis location, the "scattering efficiency" of the planet for the
specified location is calculated from the percentage of particles
that, at any point during the simulation, are scattered into an orbit
whose periastron location is interior to the snow line ($q =
a_\mathrm{ice} = 2.7$~AU) or Earth's orbit ($q = 1.0$~AU). The data
from all simulations for both suites were then compiled for the
subsequent analysis.

\begin{figure*}
  \begin{center}
    \includegraphics[angle=270,width=17.0cm]{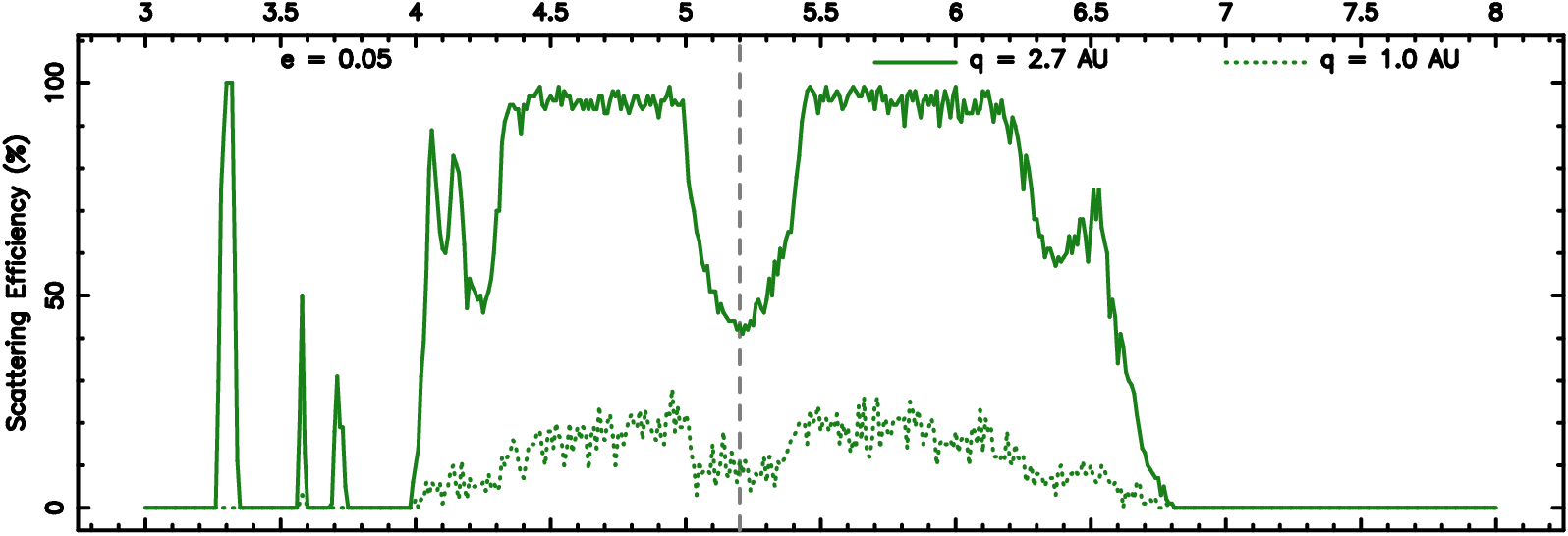} \\
    \includegraphics[angle=270,width=17.0cm]{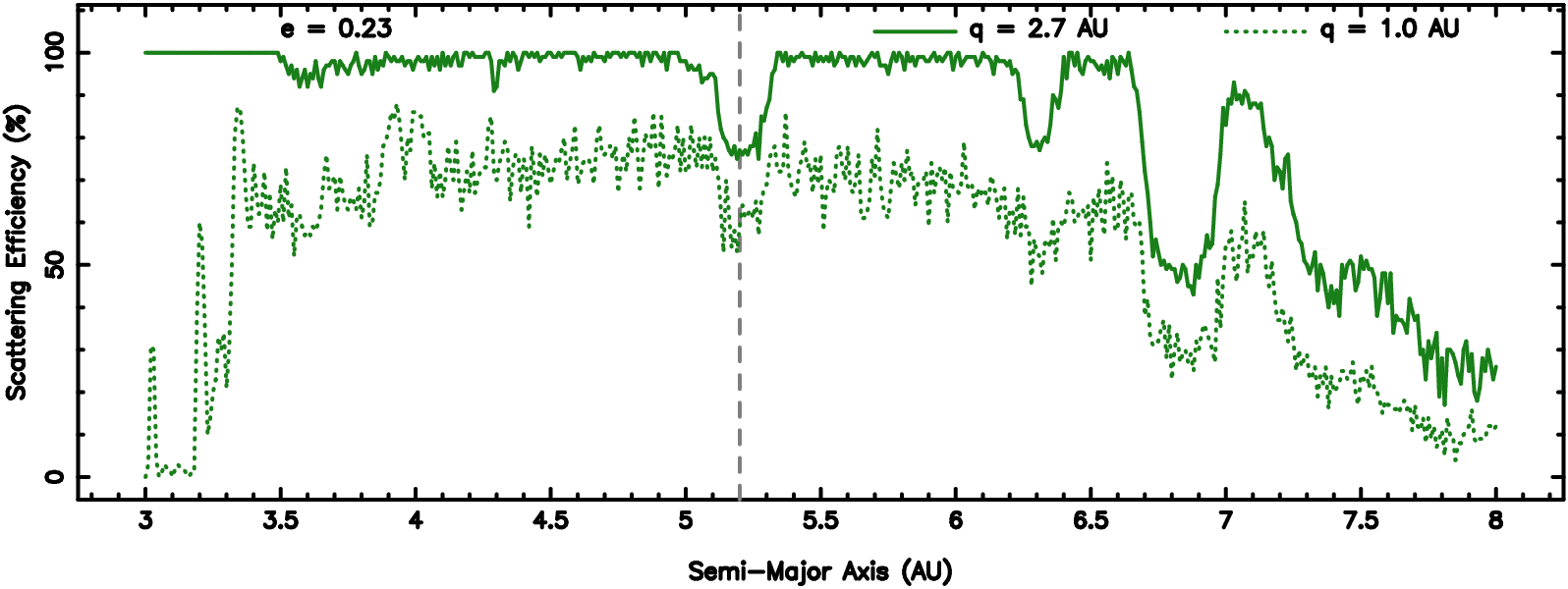}
    \end{center}
  \caption{Results for the particle injection simulation that tests
    the scattering efficiency of a Jupiter analog ($e = 0.05$; top
    panel) compared with an enhanced eccentricity Jupiter ($e = 0.23$;
    bottom panel). For each panel, results are plotted for the
    percentage of particles that achieve periastron passages interior
    to the snow line ($q = 2.7$~AU; solid line) and interior to an
    Earth orbit ($q = 1.0$~AU; dotted line). The gray vertical dashed
    line indicates the semi-major axis of the giant planet.}
  \label{fig:sim}
\end{figure*}

The particle injection simulation data are represented within the
panels shown in Figure~\ref{fig:sim}, where the scattering efficiency
results for the Jupiter analog and eccentric Jupiter cases are shown
in the top and bottom panels, respectively. The solid and dotted lines
indicate scattering interior to the snow line and Earth orbit,
respectively. The gray vertical dashed line indicates the semi-major
axis of the giant planet. For each case, and by definition, the
scattering efficiency interior to Earth's orbit is a subset of that
interior to the snow line. Both cases also show the same reduction in
scattering efficiency near the planetary semi-major axis of 5.2~AU due
to the relative stability of Trojan particles
\citep{levison1997a,morbidelli2005,nesvorny2013a,bottke2023b}. The
Jupiter analog case, shown in the top panel, clearly exhibits enhanced
scattering potential at locations of mean motion resonance (MMR),
particularly for the 2:1 ($\sim$3.2~AU), 7:4 ($\sim$3.6~AU), and 5:3
($\sim$3.7~AU) MMR locations. Locations of MMR become less efficient
at scattering when at larger semi-major axis locations due to the less
frequent gravitational perturbations from the planet. The eccentric
Jupiter case, shown in the bottom panel, produces similar MMR effects,
but the scattering potential of the planet is so enhanced by the
change in orbital eccentricity that the MMR locations are not nearly
as obvious as they are for the Jupiter analog case.

In general, the scattering efficiency of the Jupiter analog case is
less than that for the eccentric Jupiter case for the full range of
particle locations. For the Jupiter analog case, 43.6\% and 7.0\% of
particles are scattered interior to the snow line and Earth's orbit,
respectively, when integrated over all particle locations. For the
eccentric Jupiter case, 84.5\% and 54.7\% of particles are scattered
interior to the snow line and Earth's orbit, respectively. These
simulation results lead to the conclusion that the eccentric Jupiter
scatters almost double (factor of 1.9) the amount of material interior
to the snow line, and 7.8 times the amount of material into
Earth-crossing orbits, compared with the Jupiter analog case. However,
it is worth considering that the vast majority of planetesimal
material within the ``feeding zone'' of the giant planet will have
been significantly depleted during the formation of the planet
\citep{pollack1996,alibert2005b}. To account for this, we repeated the
above scattering calculations, integrated over all locations but
excluding material within three Hill radii ($\pm1$~AU) of the
planetary semi-major axis \citep{raymond2017b}. For the Jupiter analog
case, the revised total scattering efficiencies are 16.7\% and 1.6\%
for the snow line and Earth's orbit, respectively. For the eccentric
Jupiter case, the revised total scattering efficiencies are 76.5\% and
43.7\% for the snow line and Earth's orbit, respectively. Insofaras
the scattered material contains volatiles, most particularly water,
then the above described perturbative effects of the giant planet will
likely provide a significantly greater opportunity for building the
water inventory of inner terrestrial planets.


\section{Discussion}
\label{discussion}

Searches for exoplanets have always been influenced by a deep human
desire to understand the context for the solar system's orbital
configuration and, in recent decades, we have learned much about the
frequency with which the universe produces planetary systems of
various architectures. As described in Section~\ref{intro}, a key
characteristic of our solar system is the presence of long-period,
cold giant planets. Long-duration RV surveys with decades of
observations have revealed that such planets are present around
$\sim$10\% of stars
\citep[e.g.,][]{cumming2008,wittenmyer2011a,wittenmyer2020b,fulton2021,bonomo2023}. In
terms of potential solar system analogs, \citet{wittenmyer2016c}
derived an occurrence rate of $6.2^{+2.8}_{-1.6}$\% for systems
containing a low-eccentricity ($e < 0.2$) Jupiter analog and no
interior giant planets. The connection between cold Jupiters and inner
super-Earths has also been explored at length. Recent results are in
general agreement that those two types of planets appear to be
correlated \citep[e.g.,][]{zhu2018c,bryan2019,rosenthal2022}, with
nearly all cold-Jupiter systems thought to also host interior
super-Earths (but see \citet{bonomo2023} for a
counter-argument). However, as our solar system contains no
super-Earths, it may be more relevant to consider the frequency of
systems containing a cold Jupiter and no super-Earths.

The astrobiological importance in understanding the occurrence of cold
Jupiters is primarily related to their relationship to the volatile
delivery and the water inventory of inner, terrestrial planets
\citep{raymond2004a,raymond2014d,ciesla2015b,raymond2017b,marov2018,venturini2020b}. Methods
are actively being developed to detect high-water-content atmospheres,
potentially indicative of a water dominated surface, from current
transmission spectroscopic data \citep{kempton2023a}. The accumulation
of such significant atmospheric water detections will provide a
statistical context to possible water delivery from beyond the snow
line, and thus a means to test the various scenarios proposed for such
delivery. Indeed, the volatile inventory of inner planets can test the
location of the snow line with respect to pebble accretion scenarios
\citep{bitsch2021b}. It is worth noting that the nature and volume of
volatile scattering that originates from giant planet perturbations
also depends on the distribution of volatiles throughout the
protoplanetary disk
\citep{pollack1996,lodders2003,arturdelavillarmois2019a,oberg2021b}. Our
scattering efficiency results may be effectively scaled to the known
or inferred distribution of volatiles within a particular planetary
system.

The results described in Section~\ref{dynamics} show that the median
exoplanetary architecture with a giant planet beyond the snow line is
considerably more efficient at scattering volatiles interior to the
snow line, particularly to terrestrial planets that may reside within
the HZ. The main implication of these results are that water-rich
worlds may be more common in those systems with eccentric cold
Jupiters, where those terrestrial planets may contain an order of
magnitude or more water than Earth. This may not necessarily be
beneficial for habitability since research by \citet{glaser2020a}
showed that the immense seafloor pressures of water worlds can
truncate geochemical cycles that transport bioessential elements, such
as phosphorus, to ocean chemistry reactions. However, a
counter-argument by \citet{kite2018b} is that geochemical cycles with
the mantle may not be necessary when the needed bioessential element
inventories are acquired via early water-rock interactions with the
planetary crust. Water world environments may even be possible for
planets that would normally be consider mini-Neptunes
\citep{madhusudhan2021}, although such planets are unlikely to be able
to retain water in a liquid state due to water vapor participation in
a runaway greenhouse \citep{innes2023}. Clearly, there remain numerous
questions regarding the nature of water worlds, including their
origin, their prevalence, and their correlation with the presence of
cold giant planetary companions.

A further consideration is that it has been suggested from
observational data that giant planets are relatively rare around M
dwarfs
\citep{endl2006b,cumming2008,johnson2010d,bonfils2013a}. Considering
that M dwarfs are exceptionally common, both intrinsically and as
exoplanet targets for follow-up observations, the relative lack of
giant planets in these systems may require that other,
accretion-related, processes dominate the establishment of a minimum
water inventory for HZ planets \citep{marty2012,sato2016b}. Some
formation models \citep{lissauer2007,menou2013,tian2015a} and
observational data \citep{rogers2023b} support the notion that M dwarf
terrestrial planets may indeed be relatively dry. Additionally, planet
formation simulations conducted by \citet{raymond2007b} found that it
may be difficult for M dwarf accretion disks to form terrestrial
planets larger than $\sim$0.3~$M_\oplus$, potentially decreasing the
probability of M dwarfs participating in habitable planet
formation. Alternatively, M dwarf planets that form beyond the snow
line and migrate inward may retain substantial water content on the
crust and in the mantle
\citep{ogihara2009,unterborn2018a,pan2022a}. Furthermore, ice-rich
planetary embryos may migrate interior to the snow line that would
otherwise have been blocked by the presence of a giant planet
\citep{izidoro2015a,bitsch2021a}. There are numerous other particular
considerations regarding planet formation around M dwarfs, including
pebble accretion and disk heating, that are required for a full
assessment of the expected water inventory
\citep{adams2005d,kennedy2007}. Given the diversity of scenarios and
associated models for the formation and evolution of terrestrial
planets around M dwarfs, further observational evidence is necessary
to establish the influence of their formation environment and the role
of giant planets (or lack thereof) in their volatile inventory
\citep{tarter2007}.

There are several caveats to note regarding the work presented in this
paper. First, it is worth addressing if there is an observational bias
in our sample selection. As noted in Section~\ref{ecc}, there is
certainly an observational bias against the detection of low-mass
planets at long periods, but the detection of Saturn-mass and above,
as was used for our sample, is largely dominated by the RV survey
duration. Those surveys described above generally have sufficient
duration and RV precision to establish completeness to the snow lines
of the stars monitored, and often considerably further. Our use of
Saturn for the planet mass criteria was chosen as a conservative
estimate to minimize such completeness issues for the long period
planet population considered. We found lowering the mass criteria
below 0.1~$M_J$ rapidly reaches the detection limit of surveys and so
does not alter the median eccentricity of 0.23 for planets beyond the
snow line. Second, the choice of $10^5$ years for each simulation
duration is sufficient to effectively estimate the majority of
scattering events at each semi-major axis location, with the exception
of the outer semi-major axis range (as noted in
Section~\ref{dynamics}). Increasing the duration of simulations would
narrow the regions of stability as more gravitational perturbations
are accounted for. Third, scattering particles interior to the snow
line, or even to an orbit that crosses that of an inner planet, does
not translate linearly into impact rates, and thus volatile delivery
for the inner planet. Though opportunities for impacts become
available with orbit crossing events, the impact rates are generally
proportional to the cross-sectional area of the planet
\citep{horner2008a}. Fourth, there are numerous potential sources for
Earth's water inventory aside from that described here
\citep{morbidelli2012a}. For example, \citet{ikoma2006a} suggest that
Earth gained substantial water from the nebula during formation via
oxidation of a hydrogen-rich
atmosphere. \citet{delsemme1992a,delsemme1992b} proposed a cometary
origin for the bulk of Earth's water, but this proposition has been
found to be inconsistent with the D/H ratio measurements obtained from
long period comets, which is about twice that for Earth's water
\citep{balsiger1995}. For material scattered by the giant planets,
\citet{martin2021e} estimated that very little of Earth's water
originated from beyond the snow line with Jupiter at its present
location. However, moving Jupiter to the snow line can produce
significant increases in terrestrial planet water delivery,
emphasizing the importance of accounting for giant planet migration
scenarios \citep{darriba2017}. Indeed, much of the scattered material
by Jupiter and Saturn may have consisted of water-rich primitive
asteroidal material (C-type asteroids) at the outer main belt, which
in turn may have been populated from a broad range of distances beyond
the snow line \citep{raymond2009c,raymond2017b}. With all of these
volatile delivery factors combined, it is difficult to say in absolute
terms how much of the terrestrial planet inventory originates from
scattering via giant planets beyond the snow line. However, our
simulations demonstrate that it is possible to calculate the relative
effects of giant planets on volatile scattering with respect to
eccentricity, and the implications derived from the exoplanet
eccentricity distribution.


\section{Conclusions}
\label{conclusions}

Giant planets are a crucial piece of the planetary habitability puzzle
since they are often (as in the case of Jupiter) the dominant
planetary mass within the system and thus their gravitational effects
profoundly influences the system evolution. These influences include
the sculpting of the planetary architecture and the distribution and
scattering of volatiles that predominantly lie beyond the system snow
line. Given that cold giant planets are relatively rare, the lack of
their influences within the majority of planetary systems requires
careful attention when placing our solar system in a similar
context. In this work, we have demonstrated that giant planets beyond
the snow line have a median eccentricity of 0.23, at least to the
limit of current RV survey durations. This median eccentricity is
considerably higher than the 0.05 eccentricity of Jupiter. Our
dynamical simulations show that, for the solar case, a median
eccentricity Jupiter is able to scatter material to the inner part of
the planetary system substantially more efficiently than the Jupiter
analog case. Specifically, the scattering efficiency is increased by
factors of 1.9 and 7.8 for material reaching interior to the snow line
and Earth-crossing orbit, respectively. This implies that most systems
with giant planets beyond the snow line may play an even more
important role in distributing volatiles to the inner terrestrial
planets than Jupiter did in our system, and indeed the prevalence of
water worlds may be greater in the presence of such cold eccentric
Jupiters as a result. This prediction may be tested through the
detection of atmospheric water vapor and bulk planetary densities that
allow the inference of volatile inventories for a wide range of
planetary architectures.


\section*{Acknowledgements}

The authors would like to thank Sean Raymond for his valuable feedback
on the manuscript. This research has made use of the NASA Exoplanet
Archive, which is operated by the California Institute of Technology,
under contract with the National Aeronautics and Space Administration
under the Exoplanet Exploration Program. This research has also made
use of the Habitable Zone Gallery at hzgallery.org. The results
reported herein benefited from collaborations and/or information
exchange within NASA's Nexus for Exoplanet System Science (NExSS)
research coordination network sponsored by NASA's Science Mission
Directorate.


\software{Mercury \citep{chambers1999}}




\end{document}